\documentclass[epj]{svjour}

\usepackage{graphics,eucal,amsmath}

\newcommand{\kb} {  \mathrm{k}_B }
\newcommand{\HH} { {\mathcal{H}} } 
\newcommand{\LL} { {\mathcal{L}} }
\newcommand{\MM} { {\mathcal{M}} } 
\newcommand{\WW} { {\mathcal{W}} } 
\newcommand{\KK} { {\mathcal{K}} }
\newcommand{\CC} { {\mathcal{C}} }
\newcommand{\PP} { {\mathcal{P}} }

\begin{document}

\title{A schematic model for molecular affinity and binding with Ising
  variables} 
\author{Fabrice Thalmann\inst{1}}
\date{Received: \today / Revised version: date} 
\institute{Institut Charles Sadron, Universit{\'e} de Strasbourg, CNRS
  UPR 22, 23 rue du Loess, BP~84047, F-67034 Strasbourg Cedex, France}%
\abstract{%
After discussing the relevance of statistical physics in molecular
recognition processes, we present a schematic model for
ligand-receptor association based on an Ising chain. We discuss the
possible behaviors of the affinity when the stiffness of the ligand
increases. We also consider the case of flexible receptors. A variety
of interesting behaviors is obtained, including some affinity
modulation upon bond hardening of softening.  The affinity of a ligand
for its receptor is shown to depend on the details of their rigidity
profile, and we question the possibility of encoding information in
the rigidities as well as in the shape. An exhaustive study of the
selectivity of patterns with length $n<8$ is carried on. Connection
with other spin models, in particular spin glasses is mentioned in
conclusion.  }%
\maketitle

\section{Introduction}
\label{sec:intro}

Molecular biology and biochemistry differ from chemistry and physics
by the very high specificity of the interactions and the processes
that they aim at describing. Words like \textit{functions} or
\textit{shapes} are used instead of \textit{molecules},
\textit{atoms}, \textit{forces} or \textit{fields}. These biomolecules
are perceived as capable of processing information (mutual
recognition) and performing dedicated actions (switches or catalytic
reactions). It takes only two orders of magnitude in size, from 0.1~nm
to 10~nm, to abandon a world of thermal chaos and vibrations, and to
enter a world of dedicated agents and reliable
procedures~\cite{Book_Alberts_Walter}.

It was suggested by E.~Fisher that selective interaction and binding
between molecules were primarily a consequence of their complementary
shapes, which has since been known as the lock and key paradigm. These
specific interactions between highly complementary moieties accounts
for the specialized and efficient action of enzymes and often explains
how drugs work at the molecular level. The lock and key paradigm was
then extended to include final conformational changes that may happen
upon binding, a mechanism known as induced
fit~\cite{1958_Koshland}. Since a receptor and its ligand have two
complementary shapes, any structural change results in decreasing the
binding properties of the pair.

The tremendous increase of known 3d molecular structures (NMR and
X-ray scattering)~\cite{2000_Berman_Bourne,2003_Berman_Nakamura} and
the ever-growing computing capacities has made of the search for
complementary ligand-receptor interaction a very intense and
competitive research field. Numerical approaches ba\-sed on the lock and
key principle are commonly known as docking algorithms. The issue of
these strategies depend on how successfully are Coulombic forces,
hydrogen bonding, solvation properties, electronic densities\ldots
accounted for~\cite{Book_Israelachvili,2001_Leckband_Israelachvili}.

As the size of the interacting bodies increases, it is natural to
question whether statistical physics still plays a role in these
highly optimized recognition processes. In cells, most molecular
interactions are subtly balanced in order to achieve reversibility and
to prevent the occurrence of irreversible aggregation. Entropic
contributions may help to achieve this balance.

One of the current goal of computational biochemistry, still out of
reach, is to predict realistic ligand binding stabilization energies
with an accuracy of about 1~kcal.mol$^{-1}$ (1.6~$k_BT$), compared
with experimental measurements~\cite{2007_Gilson_Zhou}.  The quenching
of conformational degrees of freedom upon binding, and the subsequent
entropic changes must be accounted for when computing thermodynamical
association constants with such accuracy.

The immune system provides unrivaled cases of specific mutual recognition. 
For instance, antibodies are specialized proteins which
recognize and bind in an extremely specific and accurate manner to
foreign bodies invading a living organism. A recent bioinformatic
study of interactions between T-cells and the major histocompatibility
complex (MHC) supports the view that selective interactions between
peptides may owe more to a delicate balance among many weak additive
interactions, rather than to a strong complementary and exclusive mutual
binding~\cite{2008_Kosmrlj_Chakraborty}.

Also striking is the phenomenon of allostery. Some proteins have their
function subordinated to the presence of an effector. An historically
famous example is the transcription regulation of \textit{Lac}-operon,
for which it was demonstrated by Monod and Jacob that the fabrication
of the lactose digesting enzymes in \textit{e.coli} was conditioned to
the presence of a significant amount of lactose in the environment of
the bacteria~\cite{2005_Lewis}.  Modern biology teaches us that in the
absence of lactose, the protein \textit{Lac-}repressor binds to a
stretch of dna, and prevents the expression of the genes
downstream. When lactose molecules bind to \textit{Lac-}repressor, the
protein shape changes, and it looses its ability of binding dna,
enabling the expression of the genes under its control. The dna and
lactose binding sites are located on distinct regions of the repressor
protein, suggesting an \textit{action at distance} caused by the presence
of lactose molecules. Recently, the idea of a simple conformational
change of allosteric molecules has been challenged.  By studying a
schematic mechanical model of \textit{Lac}-repressor, R.~Hawkins and
T.C.B~MacLeish estimated the contribution of internal, vibrational
degrees of freedom, \textit{i.e.} a change in protein stiffness
induced by the ligand~\cite{2004_Hawkins_McLeish}. They concluded that
positive or negative binding entropy changes $\Delta\Delta S$ were
associated to changes (hardening or softening) in the effective spring
constants used in their mechanical model of repressor proteins. It is
precisely this idea of stiffness and entropic modulation of the
binding site efficiency that motivates the present work.

There is, as a matter of fact, a deep and formal connection between
statistical physics, recognition, binding, and information
theory. When a ligand binds selectively to a patterned substrate, it
accomplishes some kind of \textit{decoding} and reads a piece of
information conveyed by its target, the more conspicuous case being
the association of complementary base pairs in dna-dna or dna-rna
duplexes which is the cornerstone of genetic information processing.
In the following Sections, we intend to show that a simple Ising spin
chain can be turned into an elementary model for the binding of a
ligand molecule onto a patterned receptor, in the limit where both
thermal fluctuations and internal entropic degrees of freedom are
relevant. Within this framework, one can tune the interactions between
ligand and binding site, as well as the internal stiffness of the
ligand, and we also consider flexible binding sites. The binary nature
of the receptors makes them natural information carriers.  After
defining the affinity and selectivity of a ligand for its receptor in
this situation, we discuss how much dependent are the affinities and
selectivities on the stiffness parameters.  Then, we proceed by giving
examples of non monotonic behaviors of the affinities with increasing
stiffnesses. We exhibit a case of decreased affinity upon local
stiffening of the ligand, reminiscent from Hawkins and McLeish
results. We show that there are affinity biases between receptors with
similar shapes but different local rigidities.  We finally perform an
exhaustive comparison of all pairs of patterns up to a length of 8
monomers, and discuss their intrinsic ability to reliably encode
information, which is found to decrease with their length. In the
following sections, the word receptor will be used with the same
meaning as ``binding site'', \textit{i.e.} an object the size of the
ligand that binds to it.  This linguistic shortcut must not occult the
fact that in many realistic cases, the receptor is a much bigger
object than the ligand, and the binding site only constitutes a
subpart of it. This schematic model does not pretend to accurately
describe a realistic experimental situation. However, despite its
simplicity, it already displays a fairly rich phenomenology which may
find a counterpart in some real cases.

Using Ising-like models for modeling selective binding is not a new
proposal. Indeed, Schmid, Behringer and coworkers introduced and
performed intensive statistical studies of models with very similar
Hamiltonians~\cite{2006_Behringer_Schmid,2007_Behringer_Schmid,%
2008_Behringer_Schmid}.  Their model describe the contact between
hydrophobic and polar patches belonging to too opposite and
complementary binding sites, and is presented as a coarse-grained
approach to hydrophilic--hydro\-phobic interactions that are central
to both protein folding and protein-protein interactions.  Their spin
variables describe the contact between the two rigid moieties and
account for short range local rearrangements of the
coarse-grained residues. A coupling $J$ between ``spin variables'' is
interpreted as a cooperativity property, while we associate ours to a
stiffness parameter. They model 2d small rectangular patches while we
discuss more schematic 1d interacting chains; our attempt to encode
information in the stiffness pattern in addition to the spatial shape
is not found in their work. The model introduced is
Section \ref{sec:affinity-selectivity} is similar to their approach, but the
general approach of Section~\ref{sec:flexible} is original.

\section{Statistical physics of surface-bound receptors binding}
\label{sec:association-constant}

Let us consider a number of ligands $\LL$ in contact with a single
binding site $\MM$ linked to a rigid surface.  A balance is reached
resulting in an equilibrium association constant $\KK^{a}_{\LL,\MM}$
for the exchange:
\begin{equation}
\LL_{\mathrm{free}} + \MM_{\mathrm{free}} \leftrightarrow
\LL\cdot\MM_{\mathrm{bound}},
\end{equation}
from which the probability $p_{\LL,\MM}$ of observing a ligand bound
to $\MM$ obeys:
\begin{equation}
\KK^{a}_{\LL,\MM}\frac{[\LL_{\mathrm{free}}]}{C^0}(1- p_{\LL\cdot\MM}) =
p_{\LL\cdot\MM},
\label{eq:complexation-equilibrium}
\end{equation}
with brackets $[.]$ denoting molar concentrations, and $C^0$ a molar
standard reference concentration, making the equilibrium association
constant dimensionless, for instance $C^0=1$mol.L$^{-3}$.  This is a
generalization of the law of mass action that one would write for the
complexation equilibrium of two species $\LL,\MM$ in solution.
\begin{equation}
\KK^{a}_{\LL,\MM}[\LL_{\mathrm{free}}].[\MM_{\mathrm{free}}] =
[\LL\cdot\MM_{\mathrm{bound}}].C^0.
\end{equation}
The equilibrium constant depends a lot on the solvation (hydration) of
$\LL$ and $\MM$, the ionic content of the solution and all the details
of close range interactions between species.

As our current goal is to emphasize the role of the internal degrees
of freedom of $\LL$, we disregard all solvent related interactions by
making a kind of ideal solution assumption, and considering that all
ligand-receptor interactions are short ranged.  This assumption on the
solvent is equivalent to saying that all bound and unbound
conformations of the ligand receptor pair have the same solvation
free-energy. The mutual affinity of such a pair comes entirely from
shape and stiffness considerations.

It becomes possible to express the equilibrium constant
$\KK^{a}_{\LL,\MM}$ as a partition function ratio~:
\begin{eqnarray}
\KK^{a}_{\LL,\MM}
&=&\frac{\mathcal{N}_Av}{8\pi^2 V^0}\frac{Z_{\LL\cdot\MM\,\mathrm{bound}}} 
      {Z_{\LL\,\mathrm{free}}Z_{\MM\,\mathrm{free}}}.
\label{eq:equilibrium-constant-K}\\
&=& \frac{\mathcal{N}_A v}{8\pi^2 V^0} \CC^{a}_{\LL,\MM}
\label{eq:equilibrium-constant-C}
\end{eqnarray}

Equation~(\ref{eq:equilibrium-constant-K}) is a particular instance of
the equilibrium association constant derived and presented as eq.~(13)
in ref.~\cite{1997_Gilson_McCammon}. $\mathcal{N}_A$ is the Avogadro
number and $V^0$ is the volume occupied by one mole in the reference
concentration state $C^0$. $Z_{\LL\,\mathrm{free}}$ stands for the
Boltzmann-Gibbs sum over all the internal conformations of $\LL$, with
fixed orientation and center of mass. $Z_{\MM\,\mathrm{free}}$ is a
similar configuration integral over the internal conformations of
$\MM$ when $\LL$ and $\MM$ are apart and not interacting. We assume
that the configuration integral of the bound complex $\LL\cdot\MM$
reduces to a product $v Z_{\LL\cdot\MM\,\mathrm{bound}}$, where
$Z_{\LL\cdot\MM\,\mathrm{bound}}$ represents the sum over all internal
configurations of $\LL$ in contact with $\MM$ (with fixed centers of
mass and orientations) and the volume $v$ corresponds to all the
positions of the center of mass of the ligand $\LL$ relative to the
receptor $\MM$ that are considered as forming a bound state. In the
context of bimolecular associations, $v$ should range up to a few
angstrom cube ($10^{-30}~\mathrm{m}^{3}$). Note that for rigid
receptors $Z_{\MM\,\mathrm{free}}$ equals~1 by construction.

The association constant is obtained by setting $p_{\LL\cdot\MM}=1/2$
in equation~(\ref{eq:complexation-equilibrium}). One can then argue
that the partition sum per free ligand
\begin{equation}
\frac{8\pi^2}{[\LL_{\mathrm{free}}]\mathcal{N}_A}
Z_{\LL\,\mathrm{free}}Z_{\MM\,\mathrm{free}} 
\end{equation}
equals the partition sum of the bound complex
\begin{equation}
v Z_{\LL\cdot\MM,\mathrm{bound}}. 
\end{equation}
In writing these expressions, we neglected the specific volume change
occurring during the binding process.  In other words, we assume that
the Gibbs $\Delta G$ and Helmholtz $\Delta F$ thermodynamic quantities
coincide.

Equation~(\ref{eq:equilibrium-constant-K}) links the thermodynamical
affinity constant that can be experimentally determined and the
configuration integrals $Z_{\LL\cdot\MM,\mathrm{bound}}$,
$Z_{\LL\mathrm{free}}$ and $Z_{\MM\mathrm{free}}$ that are the main
focus of this work. Equation~(\ref{eq:equilibrium-constant-K}) can be
also written as:
\begin{equation}
\Delta G^{0}\simeq \Delta F^{0} = \langle\Delta U\rangle -T\Delta S_{\mathrm{c}},
\label{eq:connexion-GilsonZhou}
\end{equation}
with $\Delta F^{(0)}=-k_BT\ln(\KK^{a}_{\LL,\MM})$ and where
$\langle\Delta U\rangle$ designates the average enthalpic change of
moieties $\LL$ and $\MM$ upon binding, caused by their change in
conformation and mutual interaction, while $\Delta S_{\mathrm{c}}$
represents the corresponding change in configurational entropy. This
expression stands a particular case of  
\begin{equation}
\Delta G^{0} = \langle\Delta U\rangle+\langle\Delta W\rangle -T\Delta S_{\mathrm{c}},
\label{eq:complete-GilsonZhou}
\end{equation}
demonstrated in refs.~\cite{2007_Gilson_Zhou,%
  1997_Gilson_McCammon,2004_Mihailescu_Gilson}, where $\langle\Delta
W\rangle$ accounts for the contribution of an explicit solvent to the
formation of the ligand receptor pair. In aqueous solutions, both
hydrogen bonding and hydration forces originate from specific
interaction with the solvent. Strictly speaking, solvent mediated
interactions are associated with $\langle \Delta W\rangle$ in
eq.~(\ref{eq:complete-GilsonZhou}). However, it is to some extent
possible to take them into account by means of an effective
hamiltonian and treat these interactions as if they were part of the
direct interaction term $\langle \Delta U\rangle$.

We define the \textbf{affinity} of $\LL$ for $\MM$ as
$\CC^{(a)}_{\mathcal{L},\mathcal{M}}$
(eq.~\ref{eq:equilibrium-constant-C}). For a given number of ligands
$\LL$ and binding sites $\MM$, the affinity controls the fraction of
bound pairs of molecules (adsorption isotherms) and an increased
affinity leads to an increase of bound pairs.  However, when many
patterns $\MM_1$, $\MM_2$\ldots compete for the same ligand, the
affinity is not a good assessment of how exclusive is the
binding of $\LL$ for a given $\MM$. This is why one needs a
\textbf{relative selectivity} parameter for comparing the respective
behaviors of a ligand for a target (matching pattern $\MM$) and for a
decoy (mismatching pattern $\WW$):
\begin{equation}
\mathcal{S}_r(\LL,\WW)= \frac{ Z_{\LL\cdot\MM\,\mathrm{bound}} }
 {Z_{\LL\cdot\WW\,\mathrm{bound}}}\cdot
\frac{Z_{\WW\,\mathrm{free}}}{Z_{\MM\,\mathrm{free}}},
\label{eq:relative-selectivity}
\end{equation}
or equivalently $\mathcal{S}_r = \CC^{a}_{\LL,\MM}/\CC^{a}_{\LL,\WW}$.
We will be also interested in the \textbf{absolute selectivity}, when
comparing the affinity of the matching pattern with the affinity of a
complete set of decoys:
\begin{equation}
\mathcal{S}_a(\LL) =
\frac{Z_{\LL\cdot\MM\mathrm{bound}}}{Z_{\MM\mathrm{free}}}
\cdot\frac{1}
 {\left[\displaystyle\sum_{\WW\neq\MM}
     \frac{Z_{\LL\cdot\WW\mathrm{bound}}}{Z_{\WW\mathrm{free}}}\right] },
\label{eq:absolute-selectivity}
\end{equation}
if the ligand is given a choice between all the possible patterns, or
\begin{equation}
\mathcal{S}'_a(\LL) = \inf_{\WW\neq \MM} \left\lbrace
\frac{Z_{\LL\cdot\MM\mathrm{bound}}}{Z_{\MM\mathrm{free}}}
\cdot\frac{Z_{\WW\mathrm{free}}}{Z_{\LL\cdot\WW\mathrm{bound}}} \right\rbrace,
\label{eq:alternate-selectivity}
\end{equation}
if only one decoy $\WW$ is present at a time. The inf operator of
eq.~(\ref{eq:alternate-selectivity}) runs over all the possible
patterns $\WW$ that are different from the matching pattern $\MM$, and
picks up the best competitor of $\MM$, \textit{i.e.} the one for which
the affinity difference is the
lowest. Eq.~(\ref{eq:absolute-selectivity}) is better suited for
assessing the affinity of a ligand for its matching pattern, while all
the other possible competitors $\WW\neq\MM$ are simultaneously
present.  This definition of the selectivity depends crucially on the
set $\PP$ of allowed patterns $\WW$. Changing this set $\PP$ results
in changing the selectivity parameters $\mathcal{S}_{a,r}$. Using a
subset $\PP'$ of $\PP$ naturally leads to higher selectivities.  In
practice, a large number of poor affinity decoys can eventually beat a
good ligand-receptor pair~\cite{1996_Janin}.

\section{A minimal model for flexible ligand and rigid receptor}
\label{sec:affinity-selectivity}

We now introduce a model containing the basic components of the
above discussion: matching, internal degrees of freedom, stiffness,
information, and we search for phenomenon such as stiffness
dependent affinity and matching-decoding.

The ligand $\LL$ is modeled as an articulated chain of $n$
monomers. Each bead $i$ is allowed to occupy only two positions
labelled by a binary variable $s_i = \pm 1$ and the ligand can adopt
$2^n$ distinct internal conformations.  Ligand stiffness is enforced
by means of next nearest neighbors couplings $J_{i,i+1} = \pm J$, in
the spirit of the original spin-glass model by Edwards and
Anderson~\cite{EdwAnd,MezParVir}. 

The signs of the couplings $\lbrace J_{i,j} \rbrace$ define the ground
state shape (native shape) of $\LL$, up to a trivial two-fold
degeneracy, while the moduli $J=|J_{i,j}|$ describe the energetic cost
associated to bending distortion of the ligand (strain). We assume that
ligands have a well defined orientation, say from left to right, and we
do not consider the possibility of left-right reversal.  This can be
justified considering that biopolymers (proteins, nucleic acids)
always display such an orientation along the chain.  Finally we also
disregard the possibility of lateral shift between ligand and
receptors, as occurring for instance in the hybridization of dna
oligomers. This physically relevant situation increases significantly
the combinatorics of the association and efficient algorithms have
been designed to tackle these alignment problems~\cite{Book_Durbin}.
In this work we purposely focus more on the thermodynamical stability
of two molecules that are prepositioned in the right conformation, or
which possess a unique, non-degenerated optimal relative conformation.
The outcome of this schematic model is expected to be relevant on
length scales of about~1~nm between consecutive ``monomers'', large
enough to invoke some coarse-graining of the underlying molecules, but
small enough to preserve the importance of entropic, conformational
degrees of freedom.

\begin{figure}
\begin{center}

\resizebox{0.90\columnwidth}{!}{\includegraphics*{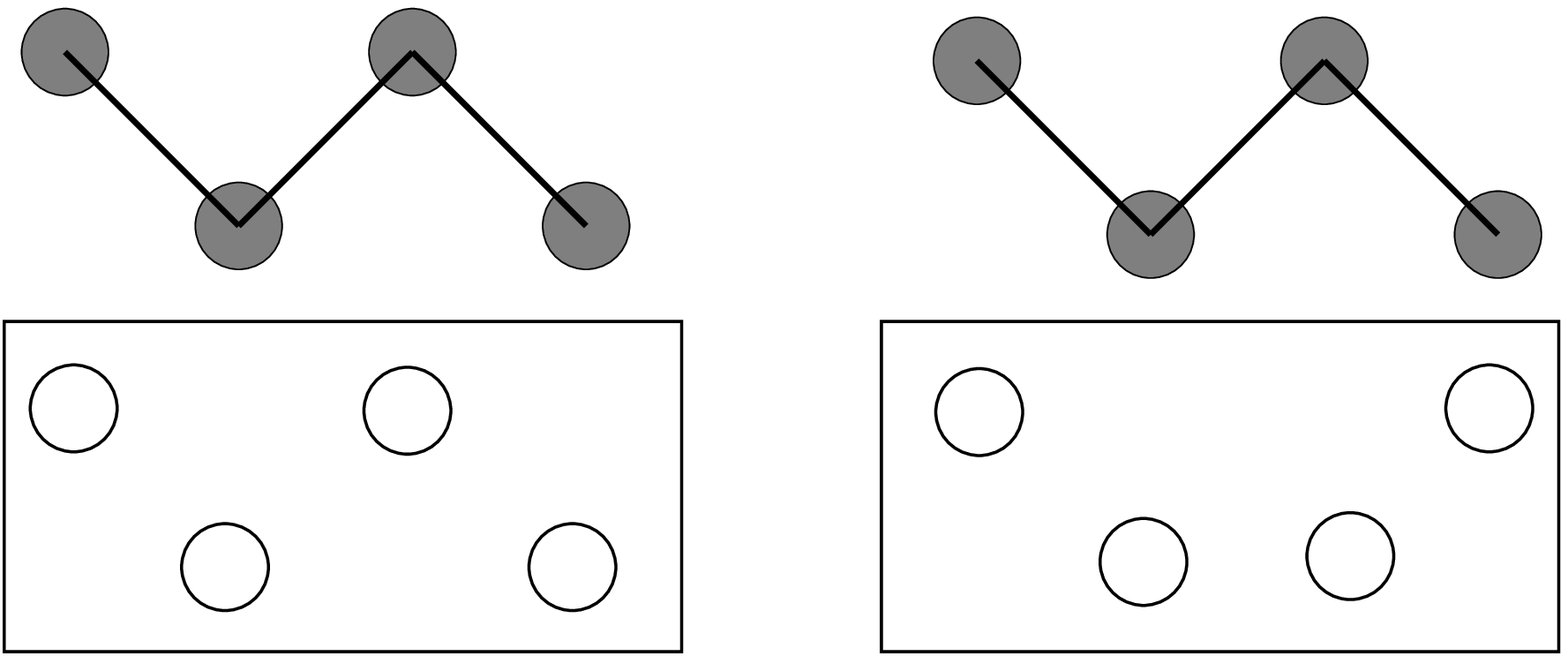} }
\caption{Picture of the ground state of a ligand molecule
  $\LL=$\texttt{u/ppp(o)} and a matching receptor
  $\MM=$\texttt{u/ppp}. With 4 monomers, the flexible ligand explores
  $2^4=16$ configuration states, while the rigid receptor does not
  change its conformation. On the right, a mismatching pattern
  $\WW=$\texttt{u/pmp} is shown.}
\label{fig:rigid-ligand-receptor}

\end{center}
\end{figure}

We now introduce a symbolic notation to describe the ground state of
these flexible molecules. For that purpose, one must distinguish
between \textit{open}, free end molecules and \textit{cyclic}, closed
end molecules. Open ligands with $n$ monomers have $2^n$ internal
configurations. There are $2^{n-1}$ possible ground states, each one
of them being doubly degenerated due to up-down symmetry. To fully
describe the shape and ground state of a ligand, we denote by the
letters \texttt{u} and \texttt{d} the position, respectively up or
down, of the first monomer on the left. We then associate a \texttt{p}
to each antiferromagnetic coupling constant $J>0$, and a \texttt{m} to
each ferromagnetic one $J<0$. A symbol \texttt{(o)} is added at the
end to signal an open chain. The ligand $\LL$ represented on the left
of Figure~\ref{fig:rigid-ligand-receptor} is thus ascribed to the
symbol $\LL=$\texttt{u/ppp(o)}. The corresponding Ising configuration
reads $s_1 = +1$, $s_2 = -1$, $s_3 = +1$, $s_4 = -1$, and the coupling
constants are $J_{12} = J_{23} = J_{34} = J$.

Cyclic ligands with $n$ monomers have spin $s_1$ and $s_n$ coupled
with a term $J_{n1}s_1s_n$. Cyclization is not really justified by the
initial molecular association problem, it is here just a convenient
way to simplify calculations and get rid of boundary effects, giving
to all monomers the same importance and simplifying the interpretation
of results. Cyclic ligands of length $n$ have the same number of
configurations as open ligands, but it requires $n$ coupling constants
to fully determine their ground state. Cyclic ligands with an odd
number of antiferromagnetic \texttt{p} couplings are ``frustrated'',
meaning that no ligand configuration can satisfy simultaneously all
the constraints imposed by the $J_{i,j}$. The ground state
conformation is then at least four-fold degenerated. Cyclic ligands
with a even number of \texttt{p} have a well defined two-fold
degenerated ground state. For instance, the unfrustrated cyclic ligand
with the same shape as represented on the left of
Figure~\ref{fig:rigid-ligand-receptor} is coded as 
$\LL'=$\texttt{u/pppp(c)}, where the suffix \texttt{(c)} reminds of
the cyclic character of the molecule. Ligand
$\LL''=$\texttt{u/pppm(c)} is frustrated with no well defined ground
shape. In this schematic approach, the use of open or cyclic ligands
is essentially a matter of convenience, as they generate qualitatively
similar results.

In the same way, the patterned receptor (binding site) is represented
by a sequence of binary values $b_i$, $1\leq i\leq n$, each one taking
a value $\pm 1$.  When the ligand comes in contact with the receptor,
it gains a negative stabilizing energy $-A$ whenever the monomer
position $s_i$ and the receptor value $b_i$ match. We do not give a
penalty to a mismatch situation $b_i\neq s_i$, but this could just be
done by shifting negatively of the total configurational energy. The
coupling constant $A$ represents a short range interaction, possibly
mimicking hydrogen bonding or hydrophobic patches. In both cases, the
effective contact parameter $A$ may depend on temperature. The
resulting total ``Hamiltonian'' describing the ligand and the receptor
in close contact is:
\begin{equation}
\HH_r\lbrace s_i \rbrace = \sum_{i=1}^{n'} \Big( J_{i,i+1} s_i s_{i+1} \Big) 
       -A \sum_{i=1}^{n} \delta_{s_i b_i}.
\end{equation}
The sum runs until $n'=n-1$ for open chains, and $n'=n$, with coupling
$J_{n,n+1}=J_{n,1}$ for cyclic chains. As $\delta_{sb}=(1+sb)/2$ for
Ising variables, the Hamiltonian:
\begin{equation}
\HH_r\lbrace s_i \rbrace = \sum_{i=1}^{n} \Big( J_{i,i+1} s_i s_{i+1} \Big) 
       -\frac{A}{2} \sum_{i=1}^{n} \Big(s_i b_i\Big) -\frac{nA}{2}.
\end{equation}
assumes the form of a random field Ising spin glass with both quench
bond disorder $J_{ij}$ and quenched random magnetic field $-A b_i/2$.
However, contrary to usual disordered systems studies, we do not
perform here any average over the quenched random bonds, as these
couplings contain the relevant information. The related partition
function, expressed with the inverse Boltzmann factor $\beta$, is
\begin{multline}
Z_{\LL\cdot\MM\,\mathrm{bound}}=\\
\sum_{\{s_i=\pm 1\} }\exp\Bigg(\frac{n\beta A}{2}+
\sum_{i=1}^{n'}( -\beta J_{i,i+1} s_i
s_{i+1}) +\,\frac{\beta A}{2} \sum_{i=1}^{n} ( s_i b_i )\Bigg). 
\label{eq:definition-ZLM-bound}
\end{multline}
Meanwhile, $Z_{\MM\,\mathrm{free}}=1$ and $Z_{\LL,\mathrm{free}}$ is a
special instance of~(\ref{eq:definition-ZLM-bound}) with $A=0$:
\begin{eqnarray}
Z_{\LL \mathrm{free}} &=&
\sum_{\{s_i=\pm 1\} }\exp\Bigg(\sum_{i=1}^{n'}( -\beta J_{i,i+1} s_i s_{i+1}
)\Bigg), \label{eq:definition-ZL-free}\\
&=&  2^n \left\lbrack \cosh(\beta J)^n \pm \sinh(\beta J)^n \right
\rbrack,\nonumber
\end{eqnarray}
result valid for cyclic chains, with a sign~$+$ without bond
frustration, and a sign~$-$ otherwise.

Finally, from the binding free energy $\Delta F$, defined
as:
\begin{equation}
\Delta F = -\kb T\ln
 \left[\frac{ Z_{\LL\cdot\MM\,\mathrm{bound}} }
 {Z_{\LL \mathrm{free}} Z_{\MM \mathrm{free}}} \right],
\label{eq:binding-free-energy}
\end{equation}
one deduces the partial enthalpic $\Delta U$ and entropic $-T\Delta
S_c$ contributions by numerically differentiating with respect to
$\beta$.
\begin{eqnarray}
\Delta U &=& \frac{\partial (\beta \Delta F)}{\partial \beta};
\label{eq:binding-energy}\\
-T\Delta S_c &=& \Delta F-\Delta U\label{eq:binding-entropy},
\end{eqnarray}
where it is assumed that $A$ does not depend on temperature
(enthalpic contribution).  

To describe the shape of rigid receptors, it suffices, in principle,
to enumerate the values $b_i$. However, anticipating the case of
flexible receptors that will be considered in the coming section, we
use for receptors the same convention as for ligands, \textit{i.e.} a
first letter \texttt{u} or \texttt{d}, followed by a list of couplings
\texttt{p} and \texttt{m}. Receptors with matching ground state are
called $\MM$, $\WW$ being associated with those with mismatching
ground states.\\

Let us illustrate the preceding section with $\LL$=\texttt{u/pppp(c)},
$\MM$=\texttt{u/pppp} and $\WW$=\texttt{u/pmpm}.  The coupling
constants of $\LL$ are $\lbrace
J_{1,2}=J_{2,3}=J_{3,4}=J_{4,1}=J\rbrace$, the matching motif
corresponds to $b_1=b_3=1$, $b_2=b_4=-1$ and the mismatching motif to
$b_1=b_4=1$, $b_2=b_3=-1$. The calculation of the partition functions
by enumeration of the 16 configurations of $\LL$, or with a transfer
matrix method gives:
\begin{eqnarray}
Z_{\LL\,\mathrm{free}} & = & 12 + 2 e^{4\beta J} + 2 e^{-4\beta J};
\nonumber\\
Z_{\LL\MM\,\mathrm{bound}} & = & e^{4\beta J} ( 1 + e^{4\beta A}) + 
4 (e^{\beta A} + e^{2\beta A}\\
& & + e^{3\beta A}) + 2e^{-4\beta J} e^{2\beta A}; \nonumber\\
Z_{\LL\WW\,\mathrm{bound}} & = & 2 e^{4\beta J} e^{2\beta A} + 4
e^{\beta A} + 2 e^{2\beta A}\nonumber\\
& & +4 e^{3\beta A} + 2 e^{2\beta A} e^{-4\beta J} + e^{4\beta A} + 1.\nonumber
\end{eqnarray}
From now on, we assume that $\kb T$ sets the energy scale, and introduce the
dimensionless coupling constants $a = \exp(\beta A/2)$, $j = \exp(\beta J)$. 
Affinity and selectivity are rational fractions of $a$ and $j$. 

\begin{equation} 
\mathcal{\CC}^{a}_{\LL,\MM}(a,j) = \frac{(1+a^8)j^4 + 4(a^2 + a^4 + a^6) + 2 a^4
    j^{-4}}  {12 + 2j^4 + 2j^{-4}},
\label{eq:exact-quenched-match}
\end{equation}
and
\begin{equation}
\mathcal{S}_r(a,j) = \frac{(1+a^8)j^4 + 4(a^2 + a^4 + a^6) + 2
  a^4 j^{-4}}  {2a^4 j^4 + 4a^2 + 2a^4 + 4a^6 + 2a^4j^{-4} + 1 + a^8}.
\label{eq:exact-quenched-mismatch}
\end{equation}

\begin{figure}
\begin{center}

\resizebox{0.90\columnwidth}{!}{\includegraphics*{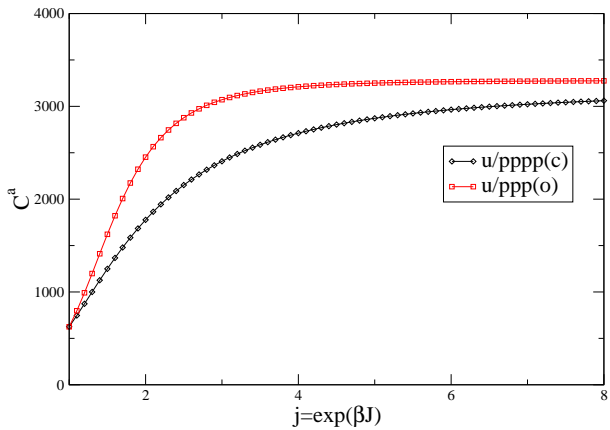}} 
\caption{Affinity $\CC^{a}$ of open $\LL=$\texttt{u/ppp(o)} and cyclic
  $\LL'=$\texttt{u/pppp(c)} with their rigid matching receptor $\MM$,
  (Figure~\protect\ref{fig:rigid-ligand-receptor} on the left) as a
  function of the ligand stiffness parameter $j$, with $a=3$. Here 
  stiffness is shown to enhance the affinity of the pair. A ligand
  whose native state matches the binding site of a receptor is said to
  be \textit{pre-organized} for matching.  }
\label{fig:affinity-rigid-match}

\end{center}
\end{figure}

\begin{figure}
\begin{center}

\resizebox{0.90\columnwidth}{!}{\includegraphics*{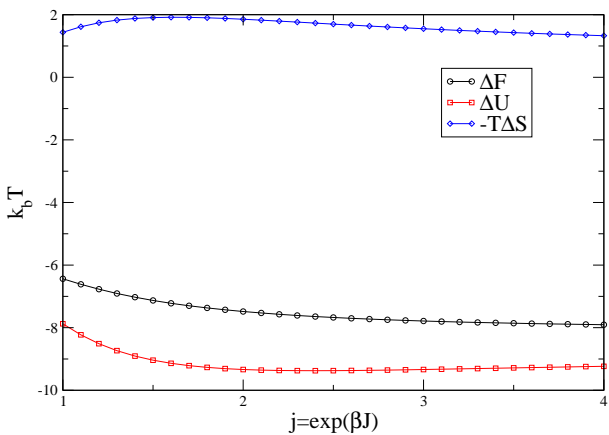}}
\caption{Thermodynamics of \texttt{u/ppp(o)} on its matching receptor,
  $a=3$, expressed in units $\beta^{-1}=kT$}
\label{fig:thermo-rigid-match}

\end{center}
\end{figure}

We are interested in assessing the role of the stiffness parameter~$j$.
In Figure~\ref{fig:affinity-rigid-match}, we observe that the affinity
of the ligand $\LL$ for the matching pattern $\MM$ increases
monotonically with $j$. Figure~\ref{fig:thermo-rigid-match} represents
the variation with $j$ of the thermodynamic potentials $\Delta F$,
$\Delta U$ and $-T\Delta S_{\mathrm{c}}$.

\begin{figure}
\begin{center}

\resizebox{0.90\columnwidth}{!}{\includegraphics*{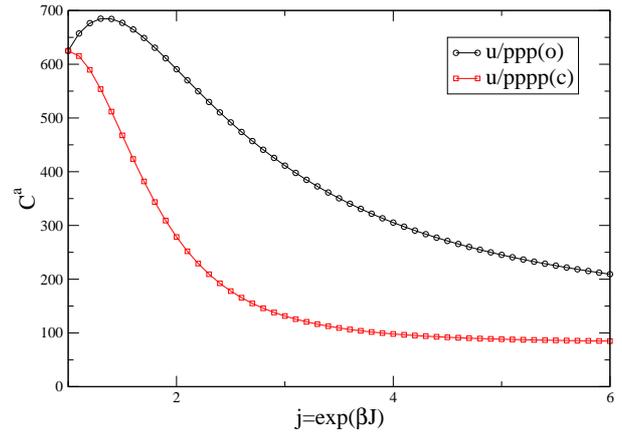}}
\caption{Affinity $\CC^{a}$ of  ligands $\LL=$\texttt{u/ppp(o)} and
  $\LL'=$\texttt{u/pppp(c)} with a mismatching
  rigid receptor (decoy $\WW=$\texttt{u/pmp},
  Figure~\protect\ref{fig:rigid-ligand-receptor} 
  on the right). The stiffness $j$ decreases the affinity of the
  cyclic pair while the affinity of the open chain decreases shortly
  after a maximum. Here $a=3$.}
\label{fig:affinity-rigid-mismatch}
\end{center}
\end{figure}

\begin{figure}
\begin{center}

\resizebox{0.90\columnwidth}{!}{\includegraphics*{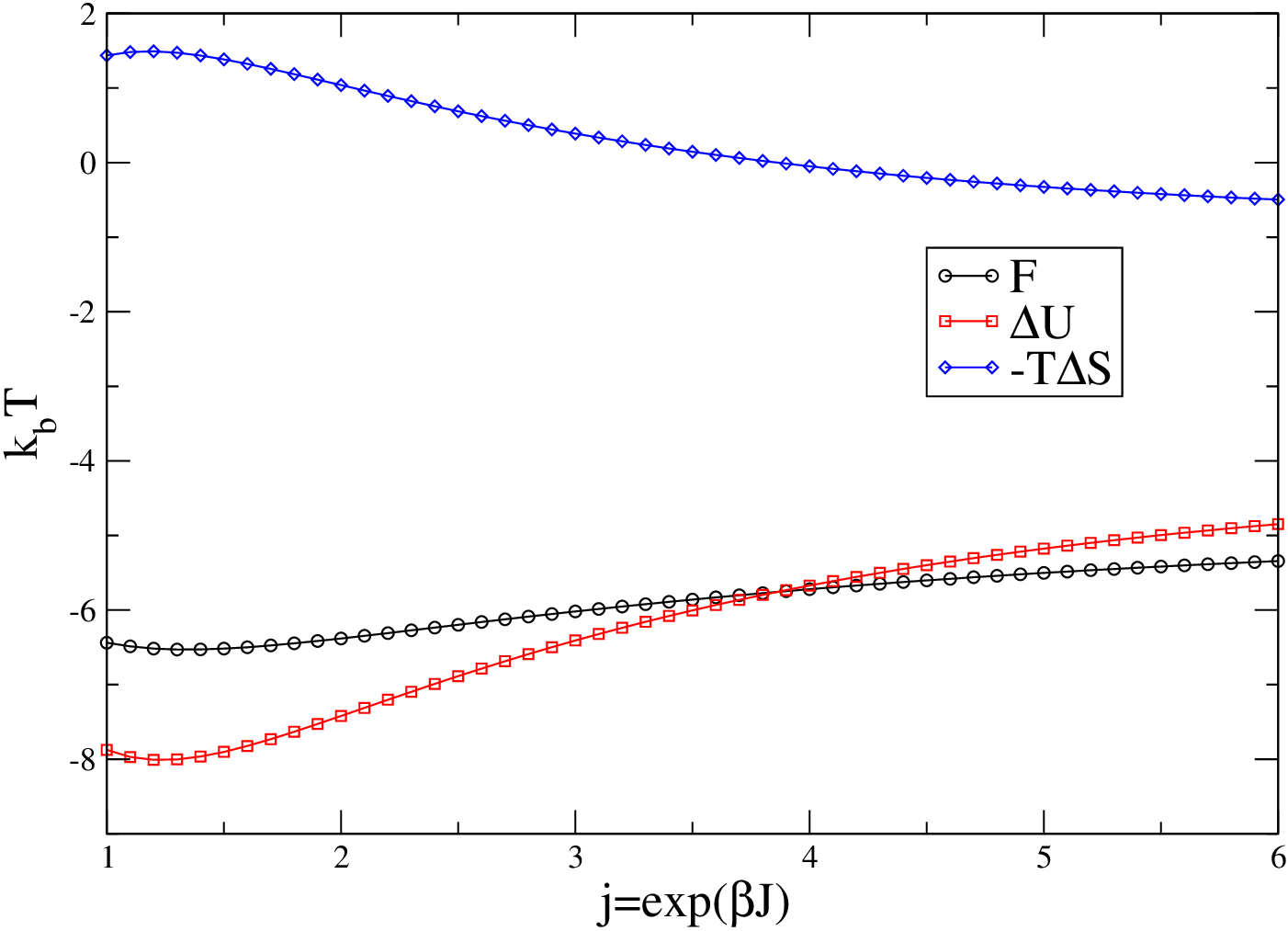}}
\caption{Thermodynamics of $\LL=$\texttt{u/ppp(o)} on receptor
$\WW=$\texttt{u/pmp}, $a=3$, expressed in units $\beta^{-1}=k_BT$.}
\label{fig:thermo-rigid-mismatch}

\end{center}
\end{figure}

At the opposite, the stiffness $j$ reduces the affinity of $\LL$ for
the mismatching pattern $\WW$, as represented in
Figure~\ref{fig:affinity-rigid-mismatch}, with the thermodynamic
functions shown in~Figure~\ref{fig:thermo-rigid-mismatch}. In
addition, one notices that the special case $j=1$ represents a soft
ligand which can adapt to any pattern. Quite naturally, the
selectivity between $\WW$ and $\MM$ is~1 for $j=1$ and tend
towards a finite value for $j\to \infty$
(Figure~\ref{fig:selectivity-rigid-mismatch}). The maximal selectivity
depends on the short range contact parameter~$a$ and is reached as
soon as $j\geq a$.

We conclude that stiffness is always favorable when the shape of a
ligand and a receptor agree, but becomes unfavorable is a mismatch is
present. When ligand and receptor shapes almost agree but not
perfectly, there must be an optimal compromise between a very soft
ligand $j=1$, which precludes any selectivity at all, and a very hard
ligand $j\gg 1$ which excessively penalizes the mismatches.

\section{Flexible receptors}
\label{sec:flexible}

\begin{figure}
\begin{center}

\resizebox{0.40\columnwidth}{!}{\includegraphics*{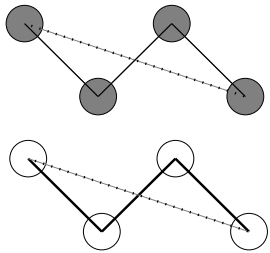} }
\caption{Ground state shape of flexible cyclic ligand
  $\LL=$\texttt{u/pppp(c)} and flexible receptor
  $\MM=$\texttt{u/pppp}. The 4~monomers ligand and receptor are
  treated on the same footing, and the partition sum encompasses
  $4^4=256$ configuration states, which is done by tracing the fourth
  power of a transfer matrix.}
\label{fig:flexible-ligand-receptor}

\end{center}
\end{figure}

The next step is to consider flexible receptors $\MM$: we expect then
soft ligands to beat stiff ligands, as they will better fit the
various configurations of $\MM$.  In our model, $\MM$ and $\LL$ play a
dual role and it becomes possible to treat ligand and binding site
(receptor) on the same footing, by inserting coupling constants $K$
between the ``spins'' $b_i$.  This situation arises when two molecules
with similar weight and structure bind together
(Figure~\ref{fig:flexible-ligand-receptor}).

\begin{multline}
\HH_r\lbrace s_i,b_i \rbrace=\\ \sum_{i=1}^{n'} \Big( J_{i,i+1} s_i
s_{i+1} \Big) -A \sum_{i=1}^{n} \delta_{s_i b_i} + \sum_{i=1}^{n'}
\Big( K_{i,i+1} b_i b_{i+1} \Big).
\end{multline}

The problem can be naturally solved for cyclic ligands and receptors
with 4x4 transfer matrices.  When all the coupling constants have same
absolute magnitude, we define $K_{i,i+1} = \eta_i |K|$, $k=\exp(\beta
|K|)$, $J_{i,i+1} = \epsilon_i |J|$, $j=\exp(\beta |J|)$ where
$\eta_i$ and $\epsilon_i$ are $\pm 1$, to find:
\begin{eqnarray}
Z_{\LL,\MM\,\mathrm{bound}} & = & a^{n}\mathrm{Tr} \left( \prod_{i=1}^{n}
T^{(\epsilon_i,\eta_i)}_{(a,j,k)} \right)\, ; \label{eq:numerical-trace}\\
Z_{\LL\,\mathrm{free}}Z_{\MM\,\mathrm{free}} & = & 4^n [\cosh(\beta J)^{n} \pm
\sinh(\beta J)^{n}]\\
& & \cdot [\cosh(\beta K)^{n} \pm \sinh(\beta K)^{n}]\, ;\nonumber
\end{eqnarray}
(the sign $\pm$ depending on the bond frustration along the chains)
with noncommuting matrices defined as:
\begin{equation}
T^{(\epsilon,\eta)}_{(a,j,k)} = \left( 
\begin{array}{cccc}
aj^{\epsilon}k^{\eta} & j^{-\epsilon}k^{\eta} & j^{\epsilon} k^{-\eta} &
a j^{-\epsilon}k^{-\eta} \\
j^{-\epsilon}k^{\eta} & a^{-1}j^{\epsilon}k^{\eta} &
a^{-1}j^{-\epsilon}k^{-\eta} & j^{\epsilon}k^{-\eta} \\
j^{\epsilon}k^{-\eta} & a^{-1}j^{-\epsilon}k^{-\eta} &
a^{-1}j^{\epsilon}k^{\eta} & j^{-\epsilon}k^{\eta} \\
aj^{-\epsilon}k^{-\eta} & j^{\epsilon}k^{-\eta} & j^{-\epsilon}k^{\eta}
& a j^{\epsilon}k^{\eta}
\end{array}
\right).
\label{eq:noncommuting-matrix}
\end{equation}

In computing the trace, one actually performs a summation over the
$4^n$ internal configurations. Thermodynamic quantities $\Delta F$,
$\Delta S_c$ and $\Delta U$, given by equations
(\ref{eq:binding-free-energy}), (\ref{eq:binding-energy}) and
(\ref{eq:binding-entropy}), are then obtained by numerically
differentiating the transfer matrix results. One notices that the
transfer matrix is entirely built from dimensionless parameters $a$,
$j$ and $k$. The numerical differentiation of $\beta\Delta F$ with
respect to $\beta$ assumes that $A$, $J$ and $K$ do not depend on
temperature, leading to $\mathrm{d} a/\mathrm{d}\beta= \beta^{-1}
a\ln(a)$, $\mathrm{d} j/\mathrm{d}\beta= \beta^{-1} j\ln(j)$ and
$\mathrm{d} k/\mathrm{d}\beta= \beta^{-1} k\ln(k)$. As a result, the
thermodynamic quantities derived from
equations~(\ref{eq:binding-free-energy}), (\ref{eq:binding-energy})
and (\ref{eq:binding-entropy}) are automatically expressed in units
$k_BT$. Non purely enthalpic contributions to the contact energy
parameter $A$ could also be included in this numerical scheme by using
a different prescription for the derivative $\mathrm{d} a/\mathrm{d}
\beta$.

The external random field $b_i$ which was quenched for rigid receptors
is now annealed (the random bonds still quenched), and we checked
that for large values of $K$ (namely $k=\exp(\beta K)\geq 5$) the
result for a flexible ligand receptor pair tends to the predictions
for the rigid receptor. One can easily convince oneself that there is
no difference between a short rigid receptor and a short flexible
receptor with doubly degenerated ground state (\textit{i.e.} no
frustration) for which the condition $k\gg j\gg 1$ holds. In this
limit, $Z_{\MM,\mathrm{free}}\simeq 2$, a factor which also appears in
$Z_{\LL.\MM,\mathrm{bound}}$, leaving $\CC^a_{\LL,\MM}$ unchanged.  In
practice, we regarded $k=10$ as sufficient to reach the rigid
situation. This corresponds to an energy gap of $3~k_BT$ between the
receptor ground state and its first distorted state. Indeed, all the
results regarding rigid receptors presented in this study were
actually obtained by setting $k$ to large values such as $k=10$ and
applying eq.~(\ref{eq:numerical-trace}).

To calculate the partition function of a flexible, open, ligand
receptor pair, one replaces the last matrix
$T^{(\epsilon,\eta)}_{(a,j,k)}$ by a matrix $T_{(a,1,1)}$ representing
the freely oscillating ends. Formula~(\ref{eq:numerical-trace})
becomes
\begin{equation}
Z_{\LL,\MM\,\mathrm{bound}} = a^{n}\mathrm{Tr} \left(T_{(a,1,1)} \prod_{i=1}^{n-1}
T^{(\epsilon_i,\eta_i)}_{(a,j,k)}\right)\,; \label{eq:numerical-trace-open}\\ 
\end{equation}

The transfer matrix formalism can be modified if one wishes to pick-up a
particular bond coupling $J_{i,i+1}$ and assign to it a value different
from the usual $J$. In our numerical implementation of the transfer matrix
product, we use a 5-letters alphabet $\{p,m,P,M,.\}$ to describe a ligand
pattern $\LL$. 
\begin{itemize}
\item Characters $p$ and $m$ are respectively used for $\epsilon = 1$
and $\epsilon=-1$. 
\item Character $.$ denotes a vanishing coupling constant $J=0$ ($j=1$).
\item Characters $P$ and $M$ represent $\epsilon = 1$
and $\epsilon=-1$, but with a different (usually larger) magnitude
of~$J$ or~$K$.
\end{itemize}

Symbols $P$ and $M$ code for some localized hardening of ligand and
receptors.  Symbol $.$ loosely connects two adjacent and stiffer
domains of $\LL$. More complex scenarios can be considered, but all
are subject to the same limitation, which is that these transfer
matrices can deal only with nearest neighbor couplings.

\section{Ligand stiffness and affinity}
\label{sec:stiffness-affinity}

We now provide a list of examples illustrating various behaviors. All
curves represent the affinity variations when the ligand stiffness is
increased from $j=1$ (soft ligand) till $j=8$, \textit{i.e.} a 2~$\kb
T$ activation barrier associated to local conformational change (spin
reversal).

\begin{figure}
\begin{center}

\resizebox{0.90\columnwidth}{!}{\includegraphics*{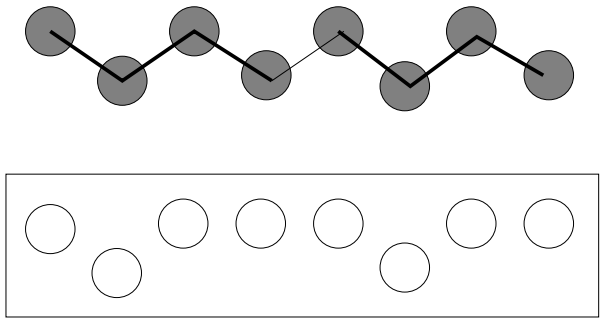} }
\caption{Ground state shape of ligand $\LL=$\texttt{u/ppp.ppp(o)} and
  motif $\MM=$\texttt{u/ppmmppm}, involved in
  Figure~(\ref{fig:p3.p3.-p2m2p2m2-k10}). The dot stands for a very
  weak coupling constant between monomers 4 and~5.}
\label{fig:M-ppp.ppp.}

\end{center}
\end{figure}

\begin{figure}
\begin{center}

\resizebox{0.90\columnwidth}{!}{\includegraphics*{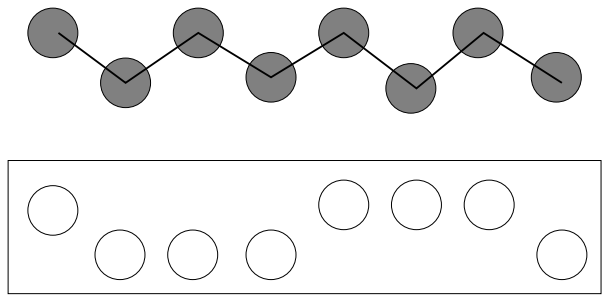} }
\caption{Ground state shape of flexible ligand
  $\LL=$\texttt{u/ppppppp(o)}, and rigid receptor
  $\WW=$\texttt{u/pmmpmmp} involved in
  Figure~(\protect\ref{fig:allp-ppmmppmm-k10b}).}
\label{fig:M-ppmmppmm}

\end{center}
\end{figure}

\begin{figure}
\begin{center}

\resizebox{0.90\columnwidth}{!}{\includegraphics*{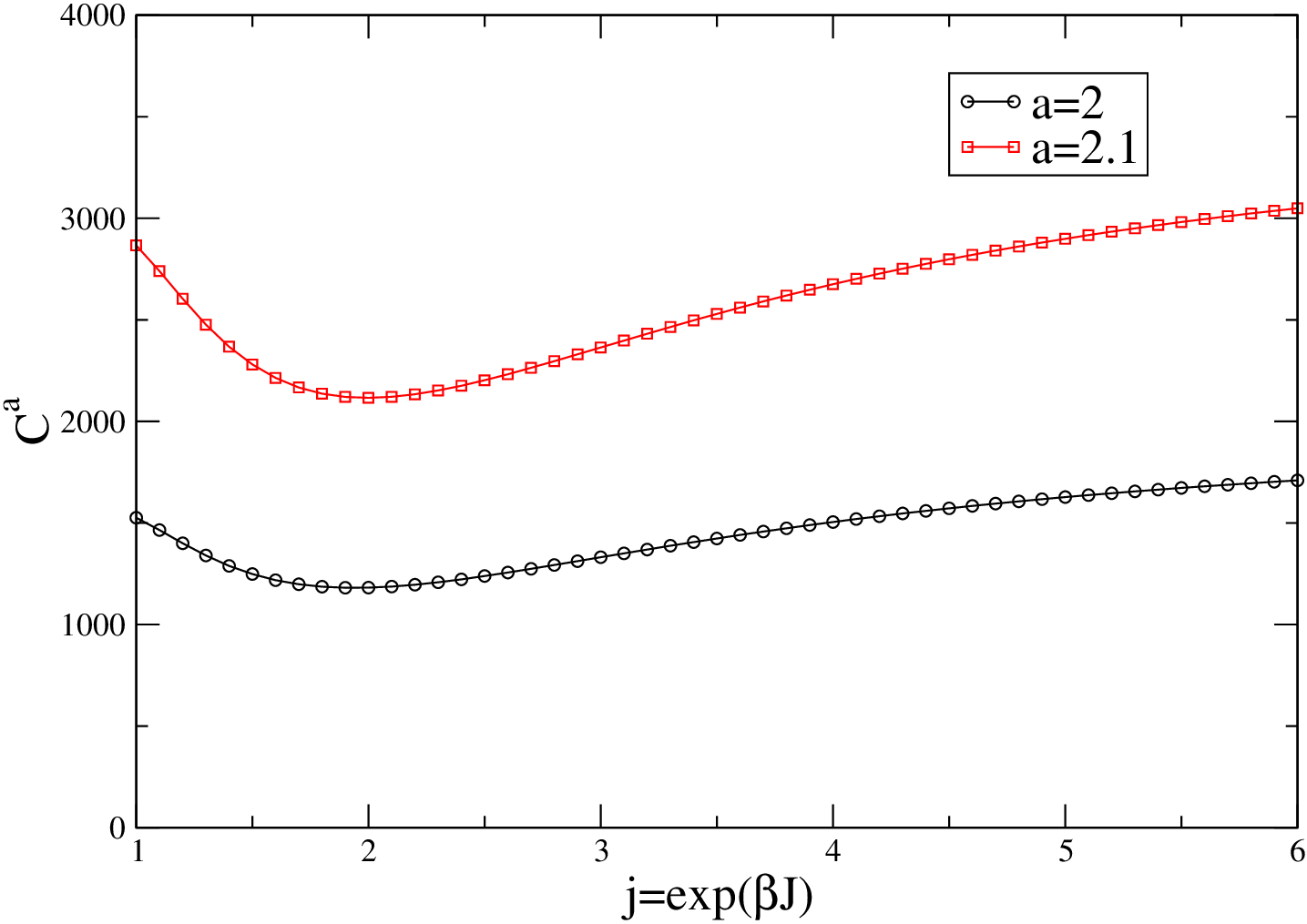}}
\caption{Nonmonotic downhill behavior of the affinity $\CC^{a}$ vs $j$
  for $\LL=$\texttt{u/ppppppp(o)}, $\MM=$\texttt{pmmpmmp}, $k=10$. The
  effect is more pronounced when $a$ increases.}
\label{fig:allp-ppmmppmm-k10b}

\end{center}
\end{figure}

\begin{figure}
\begin{center}

\resizebox{0.90\columnwidth}{!}{\includegraphics*{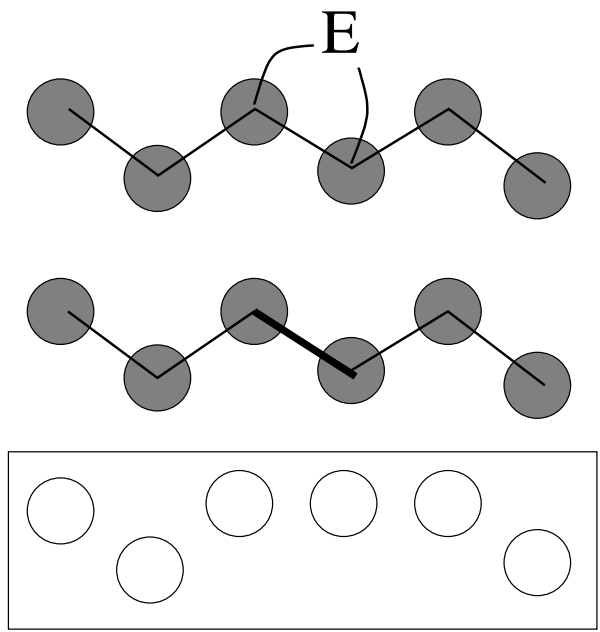} }
\caption{Ground state shape of cyclic ligand
  $\LL=$\texttt{u/ppPppp(o)} and motif $\MM=$\texttt{u/ppmmp} involved
  in Figure~\ref{fig:ppPppp-ppmmpp-a3}. The ligand possesses a hard
  bond (thick line) between monomers~3 and~4. This hard bond can be
  understood as being caused by the presence of an effector E that
  bridges the two neighboring beads~3 and~4.  }
\label{fig:M-ppPppp}

\end{center}
\end{figure}

\begin{figure}
\begin{center}
\resizebox{0.90\columnwidth}{!}{\includegraphics*{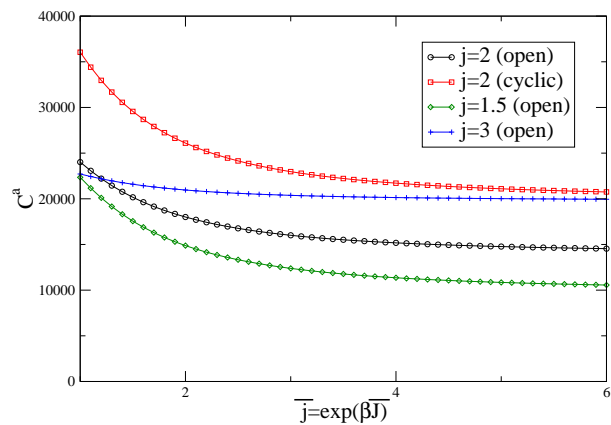} }
\caption{Affinity $\CC^{a}$ as a function of the stiffness
  $\exp(\beta\overline{J})$ of the hard link, with $j=\exp(\beta J)=2$
  and $a=\exp(\beta A/2)=3$. Because the stiff bond of the ligand lies
  on a mismatching region, its hardening leads to a decrease in
  the affinity.  Assuming that an effector molecule is the cause of
  this bond stiffening, as sketched in
  Figure~\protect\ref{fig:M-ppPppp}, one is left with a situation that
  reminds the proposed mechanism of Hawkins and McLeish.  At the
  opposite, when the stiff bond lies on a matching region, its
  hardening leads to an increase of the affinity (not shown).}
\label{fig:ppPppp-ppmmpp-a3}
\end{center}
\end{figure}

\begin{figure}
\begin{center}
\resizebox{0.90\columnwidth}{!}{\includegraphics*{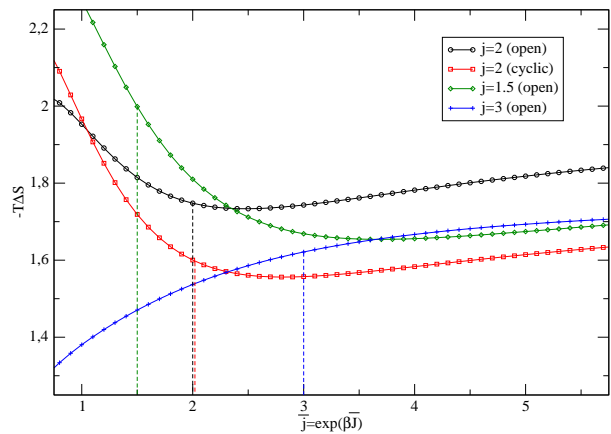} }
\caption{Configurational entropy change $-T\Delta S_c$ upon binding,
  when the stiffness of the bond linking bead 3 and 4 is increased
  (Figures~\protect\ref{fig:M-ppPppp} and
  \protect\ref{fig:ppPppp-ppmmpp-a3}). The vertical dashed lines
  indicate the points where all bond stiffnesses are identical
  $(\overline{j}=j)$. One can infer from these curves the relative
  variation of configurational entropies $\Delta\Delta S$ caused by a
  change in $\overline j$. For instance $\Delta\Delta S$ is positive
  (binding is favored) for $\overline{j}\geq j=1.5$. The trend is
  reversed for $\overline{j}\geq j=3.$, and unsettled for
  $\overline{j}\geq j=2$.}
\label{fig:DTS-uppPppx-ppmmpp}
\end{center}
\end{figure}

\begin{figure}
\begin{center}
\resizebox{0.90\columnwidth}{!}{\includegraphics*{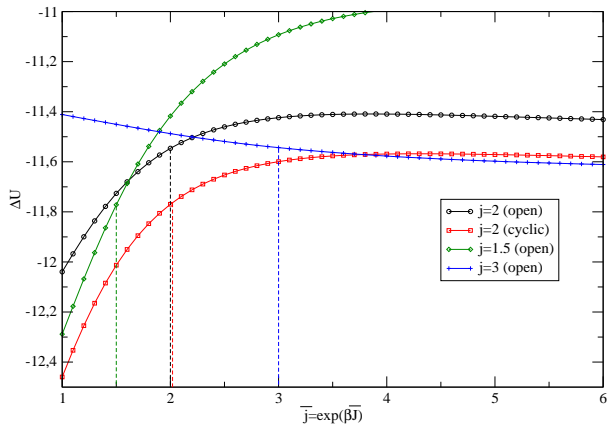} }
\caption{Energy (enthalpy) change $\Delta U$ upon binding, when the
  stiffness of the bond linking bead 3 and 4 is increased
  (Figs~\protect\ref{fig:M-ppPppp} and
  \protect\ref{fig:ppPppp-ppmmpp-a3}). The vertical dashed lines
  indicate the points where all bond stiffnesses are identical
  $(\overline{j}=j)$. The relative energy change $\Delta\Delta U$ is
  positive (binding is unfavored) for $j=1.5$ and $j=2$, but slightly
  negative for $j=3$.} 
\label{fig:DU-uppPppx-ppmmpp}
\end{center}
\end{figure}

\begin{itemize}
\item \textbf{Uphill nonmonotonic affinity}:\\ Ligand
  $\LL=$\texttt{u/ppp.ppp(o)} \textit{vs} $\MM=$\texttt{u/ppmmppm}
  (Figure~\ref{fig:M-ppp.ppp.})\\ 

The ligand $\LL$ is made of two loosely connected adjacent domains
\texttt{ppp}. The resulting affinity is nonmonotonic, first
increasing, then decreasing (Figure~\ref{fig:p3.p3.-p2m2p2m2-k10}, a
feature previously seen in
Figure~\ref{fig:affinity-rigid-mismatch}). This behavior emerges from a
competition between the matching subdomain, favored by large values of
the stiffness parameter $j$ and the mismatching subdomain whose
affinity decreases with $j$. In this particular case the mismatching
domain forces the overall affinity to decrease below its starting
point, a maximum being reached around $j\simeq 1.6$. Such a behavior
illustrates the concept of optimal stiffness, where the association of
a ligand with a (more) rigid receptor requires some tuning of its
average rigidity.\\

\item\textbf{Downhill nonmonotonic affinity}:\\
$\LL=$\texttt{u/ppppppp(o)} \textit{vs}
$\WW=$\texttt{u/ppmmppm}~(Figure~\ref{fig:M-ppmmppmm})\\ 
For $a=2$ and $k=10$ (rigid pattern), the
affinity curve is not monotonic with $j$, first decreasing, then
increasing (Figure~\ref{fig:allp-ppmmppmm-k10b}). The affinity is minimal
for a certain value $j\simeq 1.6$. This behavior in enhanced when
$a$ is increased and reduced when the receptor stiffness $k$ is
reduced, as seen in Figure~\ref{fig:allp-ppmmppmm-a2b}. The interest
of this situation is to be the exact opposite of the preceding
situation, with a local affinity minimum for $j\simeq 1.6$.\\

\item\textbf{Unbinding upon local hardening}:\\ 
Ligand $\LL=$\texttt{u/ppPpp(o)} \textit{vs} $\MM=$\texttt{u/ppmmp}
(Figure~\ref{fig:M-ppPppp})\\ 
The receptor shows a slight mismatch located under a region in $\LL$
which is locally more rigid (stiffness $\overline J > J$) than the
average coupling. Increasing $\overline J$ while keeping $J$ constant
should hamper the capacity of $\LL$ to accommodate the mismatch, and
lead to lower affinity. This is precisely what is seen in
Figure~\ref{fig:ppPppp-ppmmpp-a3}. The effect is seen for open chains
($j=1.5$, $j=2.$) and cyclic chains ($j=2$). It is reduced if the
average chain stiffness is larger ($j=3$). If one admits that the
stiff bond \texttt{P} is under control of an external agent or
effector, the issue is that the affinity of the ligand for its target
decreases even though their shape are not
altered. Figures~\ref{fig:DTS-uppPppx-ppmmpp}
and~\ref{fig:DU-uppPppx-ppmmpp} show respectively the energetic and
entropic contributions to the changes in affinity associated with
increasing the stiffness parameter $\overline{j}$. One can read from
these data the associated relative entropy change $\Delta\Delta
S=\Delta S(\overline{j})-\Delta S(j)$. Ref~\cite{2004_Hawkins_McLeish}
suggests that a positive $\Delta\Delta S$ can be associated to the
decreased affinity of \textit{lac}-repressor for dna in the presence
of lactose. This change in $\Delta\Delta S$ arises from increasing one
spring constant and decreasing another one, in an harmonic model of
\textit{lac}-repressor. In our case, because our model is not
harmonic, changes in $\Delta\Delta U$ and $\Delta\Delta S$ cannot be
separated and both contribute to changing the affinity. Note that if
the hard bond \texttt{P} was located on top of a matching subdomain,
the opposite behavior of the affinity with $\overline J$ would
occur. For instance the affinity of the pair
\texttt{u/ppPpm(o)}-\texttt{u/ppppm} increases with large values of
$\overline J$ (not shown).
\end{itemize}

To sum up, by combining matching and mismatching subdomains and bonds
of adjustable stiffness, it is possible to obtain a variety of
nonmonotonic behaviors of a ligand affinity for its target. It turns
out to be possible to control the mutual affinity by acting on the
rigidity of selected bonds, mimicking possibly the influence of a
cofactor (or effector) involved in some allosteric mechanism.

\section{Selectivity and information decoding}
\label{sec:information}

Let us consider two persons $A$ and $B$ willing to communicate, and in
possession of a number $N$ of specific ligand-receptors pairs. A sends
a ligand $i$ to B, and B brings this (still unknown to him) ligand in
contact with all the receptors available in his library. By using an
analytical tool (fluorescence, quartz crystal microbalance, resonant
surface plasmon absorption\ldots) B determines the label~$i$ of the
ligand he has received. By referring to a preestablished codebook,
shared with A, B determines the content of the message. Information
is conveyed to~B by means of selective binding~\cite{Book_Lehn}.

Information is coded in the shape of molecules, and especially
biomolecules. When the biomolecule is in solution, this information
about shape and chemical composition is made available to other
molecules and a few of them will be able to selectively bind to
it. Any selective adhesion process can be considered as information
reading, or to be more precise, information decoding.

There are a number of issues raised by this medium of
communication. One may be concerned by the reliability of the message
transmission, directly connected to the absolute selectivity
properties of the ligand-receptor pairs $\mathcal{S}_a$. One may also
asks oneself whether there are physical bounds to the minimal binding
energy or contact area necessary to ensure a transmission error rate
lower than a predefined threshold
value~\cite{Book_MaxwellDemon2,Book_Jones_Jones}.

Coding and decoding information is every day's concern for radio
engineers, and many coding schemes have been developed in order to
safely carry information around. The simplified model discussed in the
previous section establishes a convenient connection with the realm of
digital information, as ligand and receptors are already described in
terms of binary chunks of information.\\

\subsection{Dependence in the number of mismatches}

\begin{figure}
\begin{center}
\resizebox{0.90\columnwidth}{!}{\includegraphics*{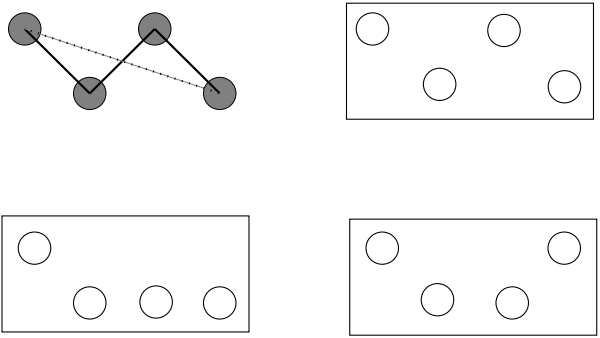} }
\caption{Ligand \texttt{u/pppp(c)} (top left) with a matching receptor
  \texttt{u/pppp}(top right), mismatching receptors \texttt{u/pmmp}
  (bottom left) and \texttt{u/pmpm} (bottom right). The Hamming
  distances in terms of configuration are respectively $d_H=0,1$ and
  2, while in terms of stiffness one has $d'_H=0$ and $d'_H=2$.}
\label{fig:M-mismatch4letters}
\end{center}
\end{figure}

I now consider the selectivity of $\LL=$\texttt{u/pppp(c)}, in contact
with altered configurations $\WW_1=$\texttt{u/pmmp} and \\
$\WW_2=$\texttt{u/pmpm}. Both altered patterns differ from the
original one respectively by one and two values of the spin $s_i$ (Hamming
distances $d_H=1$ and $d_H=2$,
\textit{cf}~Figure~\ref{fig:M-mismatch4letters}). However the Hamming
distance relative to the stiffness pattern is $d'_H=2$ for both $\WW_1$
and $\WW_2$ and is not directly related to the Hamming distance of
configurations. 

\begin{figure}
\begin{center}
\resizebox{0.90\columnwidth}{!}{\includegraphics*{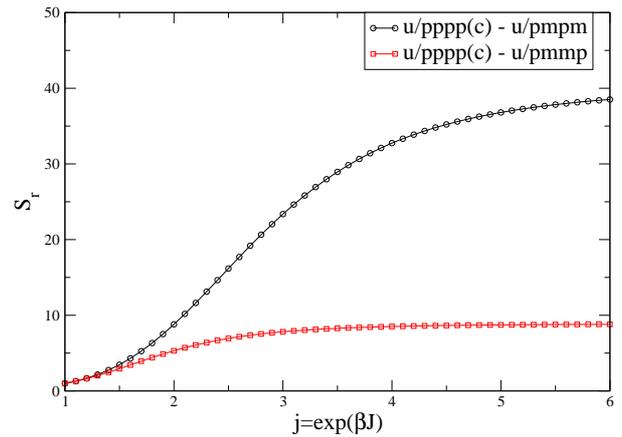}}
\caption{Selectivity $\mathcal{S}_r$ for Hamming distance $d_H=1$
  (straight lines) and $d_H=2$ (circles). The selectivity of the pair
  \texttt{u/pppp(c)}-\texttt{u/pmpm} is higher than the one of
  \texttt{u/pppp(c)}-\texttt{u/pmmp}, which means that
  \texttt{u/pppp(c)} binds \texttt{u/pmmp} ($d_H=1$) better than
  \texttt{u/pmpm} ($d_H=2$). Other parameters are $a=\exp(\beta
  A/2)=3$ and $k=\exp(\beta K)=10$.}
\label{fig:selectivity-mismatch4letters}
\end{center}
\end{figure}

Results are shown in Figure~\ref{fig:selectivity-mismatch4letters}.
It seems that the selectivity bias is indeed lower for $d_H=1$ than it
is for $d_H=2$. One expects two qualitatively different limits for the
selectivity dependence in the number of mismatches. If the stiffness
$J$ is larger than the contact energy $A$, the selectivity is likely
to correlate better with the Hamming distance $d'_H$ between stiffness
patterns: one can flip an entire interval of spins by changing only
two characters in the stiffness pattern. In the opposite limit, the
distance between configurations $d_H$ should be
dominant. Ref.~\cite{2004_Bogner_Schmid} deals in details with related
issues.\\

\subsection{Selectivity of finite length symbols}

We calculated the absolute
selectivity~(eq.~\ref{eq:absolute-selectivity}) among the subset of
all $n$-long ligand and receptor symbols, to determine the true amount
of information that such a family of ligands may carry.

To be more precise, each ground state conformation $\LL$ among the
$2^{n-1}$ possible ones (due to up-down degeneracy) was compared with
its matching counterpart $\MM$ along with all the $2^{n-1}-1$ other
competitors $\WW$, to define a shape-dependent gap of selectivity
$\mathcal{S}_a(\LL)$, eq.~(\ref{eq:absolute-selectivity}). In this
procedure, all bonds have the same stiffness $j$ and the receptors are
considered rigid, $k=10$. This shape-dependent gap of selectivity was
subsequently minimized with respect to all different possible $\LL$,
resulting in a quantity $\mathrm{Gap}(a,j,n)$ characteristic of this
family of ligands. In particular, $\mathrm{Gap}(a,j,n)$ provides a
lower bound for the occurrence of false positive if a ligand is put in
contact with an equal number of all possible receptors. Using
eq.~(\ref{eq:alternate-selectivity}) instead of
eq.~(\ref{eq:absolute-selectivity}) leads to a different quantity
$\mathrm{Gap}'(a,j,n)$ related to the noise to signal ratio when
transmitting a message using a library of compounds as suggested
earlier in this section.

Finally, the lower bound $\mathrm{Gap}(a,j,n)$ has to be optimized
with respect to the internal parameters $a$ and $j$, in order to
assess, in a context independent manner, the intrinsic coding capacity
of a family of ligands with given length $n$, defining in this way
$\mathrm{GAP}(n)$. The resulting formulas are:
\begin{eqnarray}
\mathrm{Gap}(a,j,n)&=& \inf_{\mathrm{ligands\, \LL\,of\,length}~n}
\Bigg(\mathcal{S}_a(\LL)\Bigg);\label{eq:Gap-definition}\\
\mathrm{GAP}(n) &=& \sup_{a,j}\mathrm{Gap}(a,j,n)
\label{eq:Select-definition}\\
&=&\sup_{a,j}\Bigg(\displaystyle\inf_{\mathrm{ligands\, \LL\,of\,length}~n}
\Bigg(\mathcal{S}_a(\LL)\Bigg)\Bigg).\nonumber
\end{eqnarray}
The inf operator defines the selectivity bias of a matching receptor
competing against a league of all other receptors. The sup operator
maximizes Gap with respect to $a$ and $j$.

\begin{figure}
\begin{center}
\resizebox{1.10\columnwidth}{!}{\includegraphics*{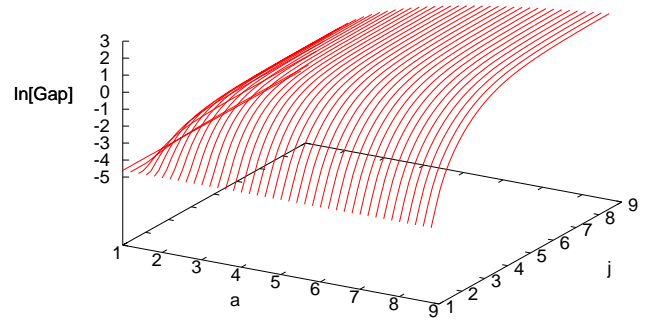} }
\caption{Three dimensional graph of the relative selectivity
  $\ln[\mathrm{Gap}(a,j,8)]$ as a function of $a$ and $j$, for
  patterns with length~8. Selectivity increases slowly with $a$ and
  $j$. Receptors are rigid, $k=10$.}
\label{fig:Selectivity-a-j}
\end{center}
\end{figure}

In practice, cyclic chains were preferred to open chains, as they
confer to all monomers the same importance.  This optimization was
performed by exhaustively scanning a rectangular grid of values
$(a,j)\in [1,9]\times[1,9]$ with step~0.1, $k$ being kept equal to 10.
Our findings are rendered as a 2d plot of Gap$(a,j,n=8)$ \textit{vs}
$(a,j)$~(Figure~\ref{fig:Selectivity-a-j}), and shows that the gap
increases with both $a$ and $j$, but grows only marginally beyond
$a,j\sim 2$. Thus, we took as representative the maximum obtained for
($a=8.8$, $j=8.8$) for estimating $\mathrm{GAP}(n)$. This arguably
constitutes only a rough estimate of the $\sup_{a,j}$ operator of
eq.~(\ref{eq:Select-definition}). Similar results were observed for
all the values of $n=3,\ldots 8$ considered in this work.  Values of
$\ln[\mathrm{GAP}(n)]$ are reported in
Table~\ref{table:Selectivities}. $\ln[\mathrm{GAP}]$ decreases
almost linearly with $n$ (Table~\ref{table:Selectivities}) as the
number of decoys grows exponentially. Extrapolating to large $n$,
GAP$(n)$ vanishes for $n\ge n^*\simeq 20$, length for which, in the
presence of an equal number of all receptors, a ligand has more chance
to bind a mismatching receptor than its own complementary one.

If one uses eq.~(\ref{eq:alternate-selectivity}) as an alternative
definition of $\mathrm{GAP}$, one observes a saturation of this
quantity to a $n$ independent value, close to~$e^{4.33}\simeq 76$ (for
$a=8.8$, $j=8.8$, $k=10$, which are quite large values). This means
that, irrespective of their length, any matching ligand receptor pair
$\LL-\MM$ has an affinity larger by a factor 76 than any other
mismatching pair $\LL-\WW$. Selectivity $\mathcal{S}'_a$ is not
sensitive to the growing number of decoys (or ``complexity'') with
$n$.

One learns from these results that if one wishes to preserve a minimal
selectivity while increasing the ligand length $n$, one must
necessarily select a subset of shapes as valid ligands and disregard
the other possibilities. This is tantamount to introducing
``redundancy'' in the ``coding scheme''. Note that the numerical value
of Table~\ref{table:Selectivities} depends on the details of the
$a$,$j$ maximization and on the choice of~$k$.\\

\subsection{Sensitivity to stiffness profiles}

The following step is to investigate the importance of stiffness
patterns compared with shape patterns. Is it possible to encode 
information in the rigidity profile~? This situation is motivated, for
instance, by the bioinformatic study of Sacquin-Mora \textit{et al.}
who investigated correlations between mechanical properties and
binding location in proteins~\cite{2007_SacquinMora_Lavery}. It would
be indeed very interesting to determine whether self assembling pairs
possess some kind of tactile sense and are sensitive to their
respective local surface rigidity.

\begin{figure}
\begin{center}
\resizebox{0.90\columnwidth}{!}{\includegraphics*{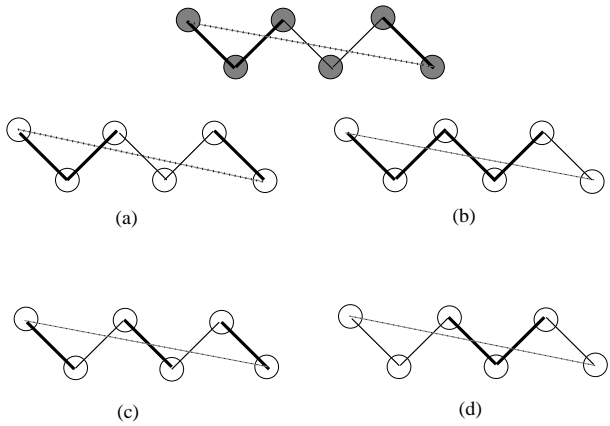}}
\caption{A ligand and four receptors with same shape but different
  rigidity patterns are represented. The ligand is
  $\LL=$\texttt{u/PPppPP(c)}, made of two rigid bonds, two soft bonds
  and then two rigid bonds (top).  The receptors are (a) middle left
  \texttt{u/PPppPP}, (b) middle right \texttt{u/PPPPpp}, (c) bottom
  left \texttt{u/PpPpPp}, (d) bottom right \texttt{u/ppPPpp}.}
\label{fig:M-PPPp-pppP-ppPp}
\end{center}
\end{figure}

We consider four identically shaped ligands and receptors with
patterns $\LL=$\texttt{u/PPppPP(c)} (ligand), \texttt{u/PPppPP},
\texttt{u/PPPPpp},\texttt{u/PpPpPp} and \texttt{u/ppPPpp} (receptors),
where \texttt{p} stands for a soft coupling and \texttt{P} for a stiff
coupling (Figure~\ref{fig:M-PPPp-pppP-ppPp}).

\begin{figure}
\begin{center}
\resizebox{0.90\columnwidth}{!}{\includegraphics*{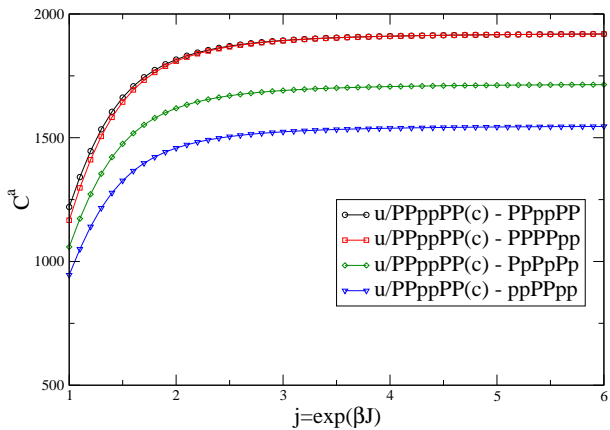}}
\caption{Affinities $\CC^{a}$ for the four pairs introduced in
  Figure~\ref{fig:M-PPPp-pppP-ppPp} as a function of $j$ (stiffness of
  the soft bonds of the ligand, symbol \texttt{p,m}), see text for
  details. Parameters are $a=2$, $k=2$, $\overline{k}=10$ (receptor
  hard bonds with symbol \texttt{P,M}), $\overline{j}=10$ (ligand hard
  bonds with symbol \texttt{P,M}).}
\label{fig:affinity-stiffness-coding}
\end{center}
\end{figure}

\begin{figure}
\begin{center}
\resizebox{0.90\columnwidth}{!}{\includegraphics*{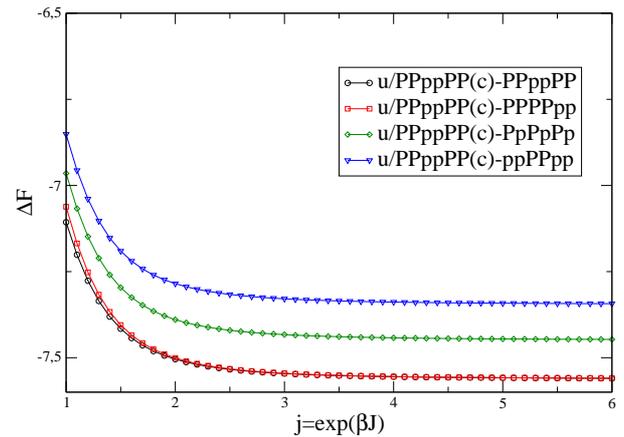}}
\caption{Free energy differences in the case of
  Figure~\ref{fig:M-PPPp-pppP-ppPp} and
  Figure~\ref{fig:affinity-stiffness-coding} result from a competition 
  between energy (enthalpy) and entropy, where the energy term is
  larger by a factor~2 compared with the entropic one. }
\label{fig:DF-PPPp-pppP-ppPp}
\end{center}
\end{figure}

\begin{figure}
\begin{center}
\resizebox{0.90\columnwidth}{!}{\includegraphics*{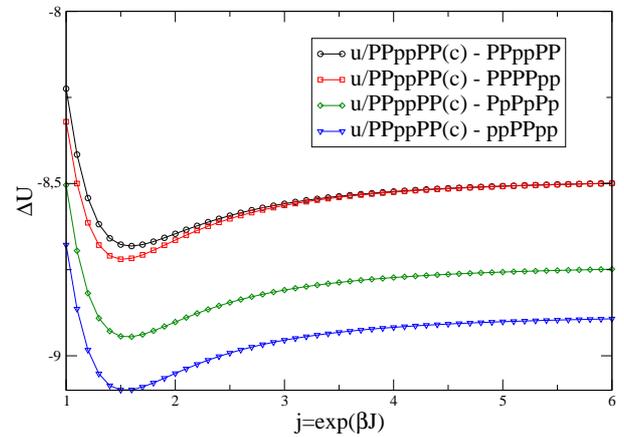}}
\caption{Energy (enthalpy) differences in the case of
  Figure~\ref{fig:M-PPPp-pppP-ppPp} and
  Figure~\ref{fig:affinity-stiffness-coding} dominates the
thermodynamics of binding.} 
\label{fig:DU-PPPp-pppP-ppPp}
\end{center}
\end{figure}

\begin{figure}
\begin{center}
\resizebox{0.90\columnwidth}{!}{\includegraphics*{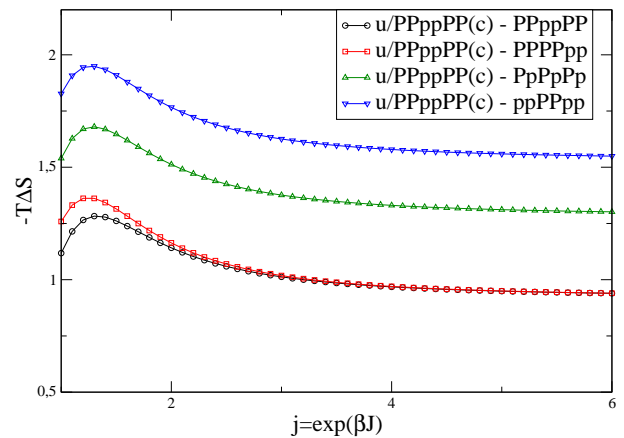}}
\caption{Differences in entropy $-T\Delta S_c$. The entropy loss is
  minimal for identical stiffness profile and maximal for opposite
  stiffness profile. }
\label{fig:DTS-PPPp-pppP-ppPp}
\end{center}
\end{figure}

The corresponding affinities are shown in
Figure~\ref{fig:affinity-stiffness-coding}. The graph shows that the
best affinity is observed when both patterns coincide (self) and the
lowest when the stiffness pattern is opposite to the ligand
(opposite). The mismatch \texttt{u/PPPPpp} is almost optimal and
the intermediate case \texttt{u/PpPpPp} gives an intermediate
value. The thermodynamic study shows that in both situations, the
entropic contribution is dominated by the enthalpic contribution
(Figures~\ref{fig:DU-PPPp-pppP-ppPp}, \ref{fig:DF-PPPp-pppP-ppPp} and
\ref{fig:DTS-PPPp-pppP-ppPp}). The stiffness patterns act on both
enthalpic and entropic terms and the issue of the competition is not
simple to predict. For instance, ligand $\LL=$\texttt{u/PpPpPp(c)} has
a stronger affinity for $\WW=$\texttt{u/PPppPP} than for its self
pattern $\MM$ (not shown).

If these results show that the affinity is sensitive to the
stiffnesses, encoding information in stiffness profiles does not seem
as straightforward as it is for shapes. Conveying information by means
of selective binding, as sketched earlier in this section, would imply
to design a quadruplet $\LL_1,\LL_2,\MM_3,\MM_4$ of ligands and
receptors with identical shapes but distinct rigidity profiles, with
affinities obeying:
\begin{eqnarray}
\CC^a_{\LL_1,\MM_1}&\sim &  \CC^a_{\LL_2,\MM_2};\nonumber\\
\CC^a_{\LL_1,\MM_2}&\sim &  \CC^a_{\LL_2,\MM_1};\nonumber\\
\CC^a_{\LL_1,\MM_1},\CC^a_{\LL_2,\MM_2} & \gg &\CC^a_{\LL_1,\MM_2},
\CC^a_{\LL_2,\MM_1}. 
\end{eqnarray}
We could not come up with such a quadruplet. These quadruplets may not
exist, or require longer lengths that the ones considered in this work
(say $n\ge 8$). Stiffness profiles modulate the recognition process,
but we were not able to prove that they could be substituted to shape
profiles. We believe it is still an open issue to know for sure if
stiffness profiles alone can bear some information.

\section{Conclusion and Perspectives}
\label{sec:conclusion-perspectives}

\begin{figure}
\begin{center}
\resizebox{0.90\columnwidth}{!}{\includegraphics*{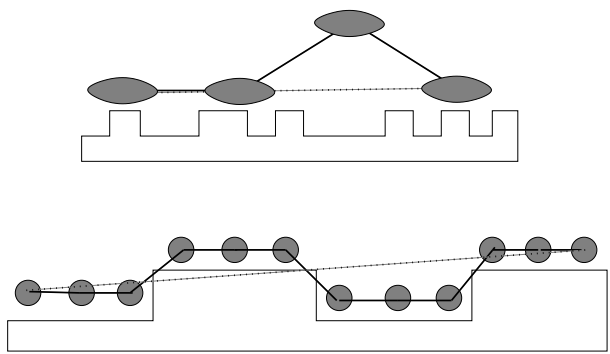}}
\caption{A coarse molecule and a fine receptor (top), a fine molecule
  and a coarse receptor (bottom).}
\label{fig:lack-freedom}
\end{center}
\end{figure}

It is clear that a ligand with too few degrees of freedom cannot
achieve good selectivity. In the top of Figure~\ref{fig:lack-freedom}, a fine
receptor is in contact with a coarse ligand, which averages out the
details of the receptor. If one changes one bit of the receptor, the
binding properties of the ligand are only marginally altered, and the
selectivity ratio stays close to~1.

At the opposite, a ligand with many monomers in contact with each
element of the receptor, will have a greater tolerance to the
fluctuations of a particular monomer (Figure~\ref{fig:lack-freedom},
bottom). Binding is enhanced by the addition of every monomer
contribution.  This case is a close analogous to the so-called
``repetition code'' in information theory. The repetition code
consists in repeating many times every single bit of information to
ensure the safe transmission of a code word (the message). It is a
greedy procedure, as the length of the message is increased by the
same factor. Here, a greater selectivity is achieved, at the expense
of a greater complexity of the molecule $\LL$ which is thrice as
long. As a rule, repetition is the easiest way one can come up with to
amplify the trends observed in Figures~\ref{fig:p3.p3.-p2m2p2m2-k10}
and~\ref{fig:allp-ppmmppmm-a2b}. By glueing together copies of the
same patterns, one automatically enhances the characteristics of the
ligand-receptor repeated unit.

To summarize a comparison with the results presented
in~\cite{2006_Behringer_Schmid,2007_Behringer_Schmid,%
2008_Behringer_Schmid}, we can say that we focussed mostly on
interesting characteristic trends exhibited by a few selected pairs of
ligand and receptors, while the authors
of~\cite{2006_Behringer_Schmid,2007_Behringer_Schmid,%
2008_Behringer_Schmid} favor global and averaged trends running over
the whole set of possible patterns. Our definition of the Gap and GAP
indicators differ from the free energy difference $\Delta F$ which
serve as a criterion in their work. Our transfer matrix approach
treats exactly the binding statistics of a given pair of ligand and
receptor, without need of Monte-Carlo sampling, nor mean-field or
large $J$ approximations that are required in their 2d approach.
Finally we believe that our model is the first one that makes it
possible to investigate the role of local stiffness modulation, and
the possibility of stiffness encoding of information.

The connection between error-correcting transmission codes and spin
systems was recognized by N.~Sourlas~\cite{1989_Sourlas,%
  2001_Sourlas,2001_Nishimori}. In particular, it was shown that the
usual binary parity checks $b_1\oplus b_2 \oplus \ldots \oplus b_p$
which involves the sum of $p$ binary digits modulo~2 was in fact
equivalent to coupling $p$ spins $s_1s_2\ldots s_p$, with $s_i=\pm 1$,
leading to a formal connection between information theory and $p$-spin
glasses.  Our model is currently restricted to nearest neighbors
coupling, and cannot account for long range couplings. Information
redundancy is thus limited to simple repetition codes. It would be
interesting to investigate how a second layer of spins and couplings
could be added in order to better enforce robustness with respect to
single bit mismatches. One also notices that a 2d generalization of
the ligand shape would indeed be equivalent to a genuine 2d
Edwards-Anderson spin glass.  Spin glasses are well known for their
long-lived or metastable states~\cite{MezParVir,1995_Monasson}. Each
one of these states can be put in correspondence with a matching
random-field representing a different receptor.

Another interesting connection between spin systems and pattern
recognition, is the \textit{Superparamagnetic clustering of
  data}~\cite{1996_Blatt_Domany,1998_Wiseman_Domany}. Wiseman, Blatt
and Domanyi showed that it was possible to train a two dimensional
array of Potts spins in order to recognize picture features and
patterns (2d inhomogeneous distributions of points). This work could
provide hints on how to train a 2d elastic network for shape and
stiffness recognition.

The nonmonotonic behaviors of the affinity, or the decrease of affinity
upon local stiffening of the chain are still modest, showing only a
cut by half in the case illustrated in
Figure~(\ref{fig:ppPppp-ppmmpp-a3}). One may be interesting in finding
stronger effects by hardening more than a single bond. 

The schematic model introduced and studied here is intended to guide
us towards more realistic examples, such as simple molecules that
could be designed and investigated with the help of \textit{coarse
  grained} or \textit{all atom} numerical models. This, we believe,
should be the next step to endeavor.\\

\section*{Acknowledgements}
\label{sec:ackowledgements}
The author thanks Carlos Marques for discussions on this topic, and is
indebted to the referees for many meaningful comments.


\clearpage


\clearpage 

\begin{figure}
\begin{center}

\resizebox{0.90\columnwidth}{!}{\includegraphics*{FIG6-Select-upppx-upmpm-a3-k10.eps}}

\caption{Selectivity $\mathcal{S}_r$ between ligand
  $\LL=$\texttt{u/ppp(o)}, $\LL'=$\texttt{u/pppp(c)}, their matching
  rigid receptor $\MM=$\texttt{u/ppp} and a mismatching receptor
  $\WW=$\texttt{u/pmp}, as a function of the ligand stiffness $j$ (see
  eq.~\protect\ref{eq:relative-selectivity}). $\mathcal{S}_r$ is
  directly related to the probability of the ligand to bind the decoy
  instead of the matching pattern. As expected $\mathcal{S}_r=1$ when
  $j=1$ because a completely soft ligand binds equally well any
  receptor (non specific binding). Here $a=3$. }
\label{fig:selectivity-rigid-mismatch}
\end{center}
\end{figure}

\begin{figure}
\begin{center}

\resizebox{0.90\columnwidth}{!}{\includegraphics*{FIG9-Affin-uppp.pppx-uppmmppmm-k10.eps} }
\caption{Nonmonotic behavior of the affinity $\CC^{a}$ vs
  $j$ for $\LL=$\texttt{u/ppp.ppp(o)}, $\MM=$\texttt{u/ppmmppm},
  $k=10$. The effect is more pronounced for larger values of $a$.}
\label{fig:p3.p3.-p2m2p2m2-k10}
\end{center}
\end{figure}

\begin{figure}
\begin{center}

\resizebox{0.90\columnwidth}{!}{\includegraphics*{FIG12-Affin-upppppppx-pmmpmmp-a2.eps} }
\caption{Nonmonotic behavior of the affinity $\CC^{a}$ vs $j$ for
  $\LL=$\texttt{u/ppppppp(o)}, $\MM=$\texttt{u/pmmpmmp}, $a=2$. The
  effect disappears when the flexibility $k$ of the receptor decreases
  and the limit case $k=1$ corresponds to a very soft, patternless
  receptor.}
\label{fig:allp-ppmmppmm-a2b}

\end{center}
\end{figure}


\clearpage 

\clearpage 

\begin{table}
\begin{center}
\begin{tabular}{||c|c|c||}
\hline
Length $n$ & $\ln(\mathrm{GAP})$\\
\hline
2 & 3.65\\
\hline
3 & 3.22\\
\hline
4 & 2.91\\
\hline
5 & 2.67\\
\hline
6 & 2.47\\
\hline
7 & 2.30\\
\hline
8 & 2.15\\
\hline
\end{tabular}
\caption{Table of selectivities for patterns with increasing sizes.
See text for details.}
\label{table:Selectivities}

\end{center}
\end{table}

\end{document}